\documentclass[11pt,aps,showpacs,showkeys,preprint,eqsecnum]{revtex4}

\usepackage{amssymb}
\usepackage{amsmath}
\usepackage{graphicx}
\usepackage{amsbsy}
\usepackage{color} 
\usepackage{ulem} 
\newcommand{\bq}{\begin{eqnarray}}
\newcommand{\eq}{\end{eqnarray}}
\newcommand{\bqn}{\begin{eqnarray*}}
\newcommand{\eqn}{\end{eqnarray*}}

\begin{document}
\title{Fluid-Fluid and Fluid-Solid transitions in the Kern-Frenkel model from Barker-Henderson thermodynamic perturbation theory}  

\author{Christoph G\"ogelein}
\email{christoph.goegelein@ds.mpg.de}
\affiliation{Max-Planck-Institute for Dynamics and Self-Organization (MPIDS), 37077 G\"ottingen, Germany}

\author{Flavio Romano}
\email{flavio.romano@gmail.com}
\affiliation{Physical and Theoretical Chemistry Laboratory, Department of Chemistry, University of Oxford, Oxford OX1 3QZ, United Kingdom}

\author{Francesco Sciortino}
\email{francesco.sciortino@uniroma1.it}
\affiliation{Dipartimento di Fisica and
  CNR-ISC,  {\it Sapienza} Universit\`a di Roma, Piazzale A. Moro 5, 00185 Roma, Italy}

\author{Achille Giacometti}
\email{achille@unive.it}
\affiliation{Dipartimento di Scienze dei Materiali e Nanosistemi, Universit\`a Ca' Foscari Venezia,
Calle Larga S. Marta DD2137, I-30123 Venezia, Italy}

\date{\today}

\begin{abstract}
We study the Kern-Frenkel model for patchy colloids using Barker-Henderson second-order thermodynamic perturbation theory. 
The model describes a fluid where hard sphere particles are decorated with one patch, so that they interact via a square-well (SW) potential if 
they are sufficiently close one another, and if  patches on each particle are properly aligned.
Both the gas-liquid and fluid-solid phase coexistences are computed and contrasted against corresponding   Monte-Carlo simulations results.
We find that the perturbation theory describes rather accurately numerical simulations all the way from a fully covered square-well 
potential down to the Janus limit (half coverage).
In the region where numerical data are not available (from Janus to hard-spheres), the method provides estimates of the location of the critical lines
that could serve as a guideline for further efficient numerical work at these low coverages.
A comparison with other techniques, such as integral equation theory, highlights the important aspect of this methodology in the present context.
\end{abstract}



\maketitle
\section{Introduction}
\label{sec:introduction}
Perturbation theory has a long and venerable history in the context of fluids and a detailed description
of several different techniques is presented in  
classic textbooks \cite{Hansen86,Gray84}, and in excellent dedicated reviews \cite{Henderson71,Barker76}.

Although the general idea dates back to a much earlier time, the first well established paradigm of first- and second-order
perturbation theory was devised by Zwanzig \cite{Zwanzig54} for simple fluids, later extended to polar fluids \cite{Zwanzig55}.
A similar analysis was carried out by Buff and Schindler in the context of solution theory \cite{Buff58}.

In addition to these theories that assume the hard-spheres model as unperturbed system, other theories exist that rely on the van der Waals 
picture as a starting point, the best known of these being the Weeks-Chandler-Anderson (WCA) theory 
\cite{Chandler70,Weeks71,Andersen76,Chandler83}. 

While the WCA theory has proven extremely powerful in many applications, for potential with hard-cores the original Zwanzig theory offers 
a natural scheme, hinging on an unambiguous potential separation. This was eventually put on firm ground by Barker and Henderson (BH)
\cite{Barker67,Henderson71,Barker76} who provided reliable estimates for square-well fluids \cite{Henderson80},
a rather unrealistic potential in the framework of simple liquids, but much more physically sound when applied to the colloid domain.

In the present paper, we will apply the BH thermodynamic perturbation theory to the Kern-Frenkel (KF) model for patchy colloids 
\cite{Kern03,Bianchi11}. In this model \cite{Kern03}, attractive circular patches are distributed on the surface of hard-spheres, and
different spheres attract each other provided that any two patches on  distinct spheres are suitably aligned, and the relative radial distance
between the centers of the spheres is within the range of the attractive tail.  

While not new \cite{Lomakin99}, these systems have witnessed an impressive resurgence of interest in the last few years for a number of reasons. 
The first reason is due to the remarkable improvements in  the chemical synthesis techniques that allows to decorate the
surface of a colloid with great precision and reliability, a feature that was not possible until few years ago.
When combined with the additional advantage, as compared with
their atomic counterpart, of an almost arbitrarily control of their size and interaction range, this makes patchy colloids very attractive
for technological applications, as elementary building blocks for self-assembly materials of the new generation \cite{Glotzer04,Glotzer07}. 
An additional important reason hinges on the fact that patchy colloids may serve as a paradigm for systems with low valence, strong anisotropy,
and highly directional interactions between particles, a feature that is common to many different systems, globular proteins being a notable
example, where the heterogeneity of the surface groups cannot be neglected even at the minimal level.

Several examples of applications and improvements of the original BH scheme have been offered over the years. Verlet and Weiss 
discussed a comparison with numerical simulations and experimental results both for simple \cite{Verlet72} and polar \cite{Verlet74}
fluids; Gubbins and Gray \cite{Gubbins72} proposed a perturbation scheme for molecular fluids; Chang and Sandler \cite{Chang94} exploited it 
to develop an analytical approximation for the square-well fluid valid within a particular interval of well amplitude; Zhang et al, 
\cite{Zhang01} applied it to a square-well chain fluid, whereas Rotemberg {\it et al.} \cite{Rotemberg04} used it to study the phase behavior 
of mixtures of colloidal particles and interacting polymers. More recently, Zhou \cite{Zhou07} derived a simple procedure hinging on the 
BH scheme to locate the fluid-solid coexistence phase for a hard-core attractive Yukawa fluid, and Kalyuzhnyi {\it et al.} 
\cite{Kalyuzhnyi10,Kalyuzhnyi11} tackled the single and multiple patchy colloids, similar to those treated in the present work, using 
a generalization of Wertheim's thermodynamic perturbation theory \cite{Wertheim84,Wertheim86a, Wertheim86b,Wertheim87}. 

The present work builds upon the methodology outlined in Ref. \cite{Goegelein08} to show that BH second-order perturbation theory
can be successfully applied to patchy colloids, as represented by the Kern-Frenkel model \cite{Kern03}. Besides thermodynamic quantities,
such as virial equation of state and chemical potentials, the method allows a rather precise location of the fluid-fluid and the fluid-solid
coexistence lines, in principle for arbitrary number and size of the patches. In this respect, the method competes in accuracy with integral equation
theory on the same system \cite{Giacometti09b,Giacometti10}, without suffering from the unavoidable instabilities present in that
case for low surface coverages and temperatures. This will be demonstrated by an explicit comparison with numerical simulations carried out
\cite{Sciortino09,Sciortino10,Giacometti09b,Giacometti10} on the same system.

The outline of the paper is as follows. In Section \ref{sec:model} the model is defined and in Section \ref{sec:perturbation} the used perturbation
technique is described. Some technical details of the calculations are included in Appendices \ref{sec:appa} and \ref{sec:appb}. Section
\ref{sec:coexistence} includes the method to compute the coexistence curves from the analytical results, with details of the numerical procedure
included in Appendix \ref{sec:appc}. Section \ref{sec:montecarlo} briefly summarize some details of the Monte Carlo calculations,
and Section \ref{sec:results} includes descriptions of all results. Finally, Section \ref{sec:conclusions} summarizes the paper
and provides some future perspectives.
\section{The Kern-Frenkel model}
\label{sec:model}
\begin{figure}[htbp]
\begin{center}
\vskip0.5cm
\includegraphics[width=10cm]{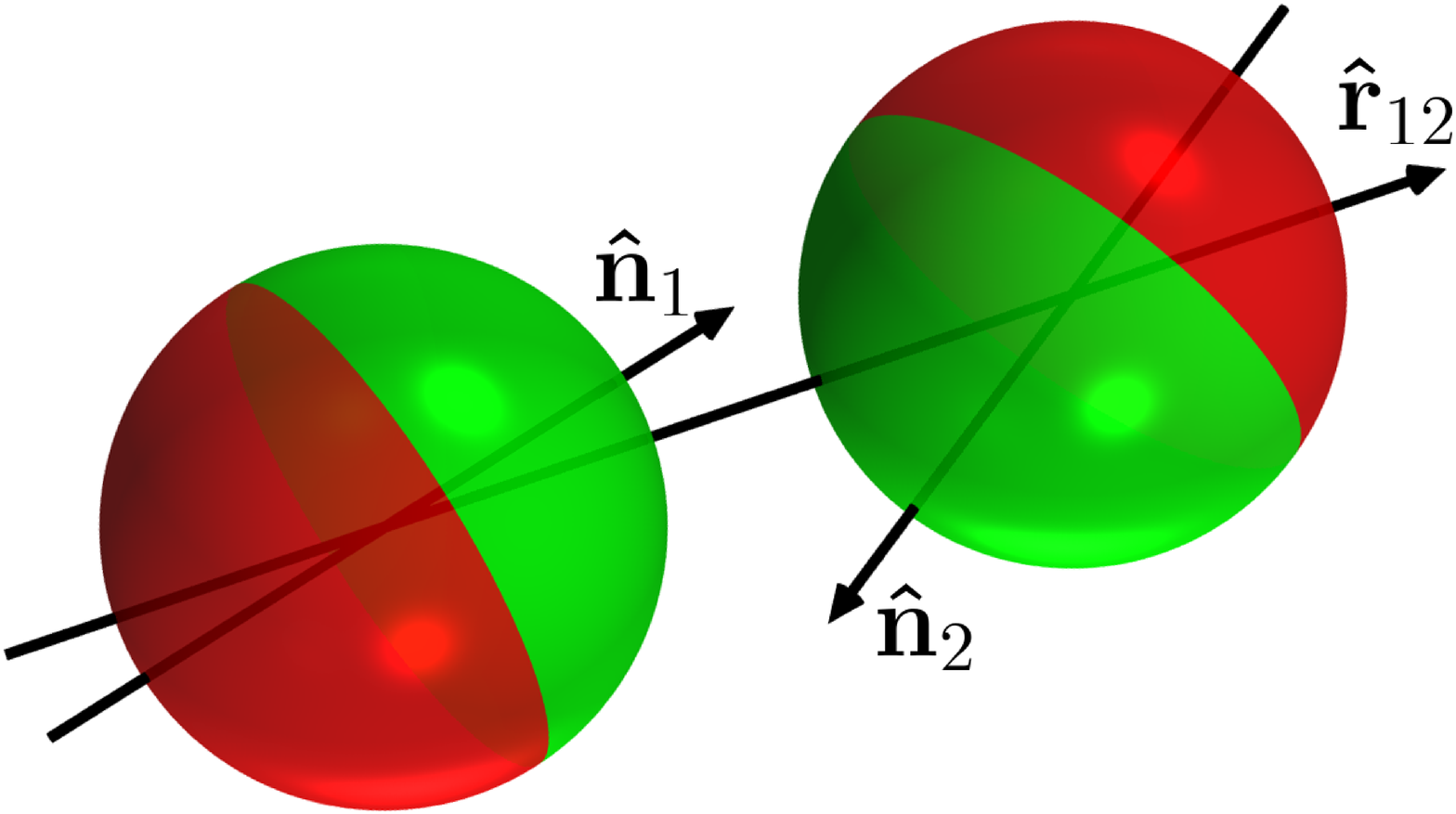}
\caption{The Kern-Frenkel potential in the case of a single patch. The surface of each sphere is partionated into an attractive
part (color code: green) and a repulsive part (color code: red). 
Units vectors $\hat{\mathbf{n}}_{1}$ and $\hat{\mathbf{n}}_{2}$ identify the directions
of each patch, whereas the unit vector $\hat{\mathbf{r}}_{12}$ join the centers of the two spheres, directed from sphere $1$ to sphere $2$. 
The particular case shown corresponds to a $50\%$ fraction of attractive surface (coverage $\chi=0.5$).  
\label{fig:fig1}}
\end{center}
\end{figure}

Consider a fluid formed by $N$ particles in a volume $V$ at temperature $T$, and assume that they can be described by the Kern-Frenkel model 
\cite{Kern03} in its one-patch version (see Fig. \ref{fig:fig1}), 
where the orientation of the patch on each surface sphere 
$1$ and $2$ is identified by unit vectors  $\hat{\textbf{n}}_{1}$ and  $\hat{\textbf{n}}_{2}$, whereas the direction connecting centers of
spheres $1$ and $2$ are identified by unit vector $\hat{\mathbf{r}}_{12}$.

Two spheres of diameter $\sigma$ attract via a square-well potential of width $(\lambda-1) \sigma$ and depth $\epsilon$ if the directions 
of the patches on each sphere
are within a solid angle defined by $\theta_0$, and repel each other as hard spheres otherwise. The pair potential has the form  
\begin{eqnarray}
\Phi\left(12\right) &=& \phi_{0} \left(r_{12}\right) +\Phi_{\text{I}}\left(12\right)\,\mbox{,}
\label{kf:eq1}
\end{eqnarray}
where the first term is the hard-sphere contribution
\begin{equation}
\phi_{0}\left(r\right)= \left\{ 
\begin{array}{ccc}
\infty,    &  &   0<r< \sigma    \\ 
0,          &  &   \sigma < r \ %
\end{array}%
\right.\,\mbox{,} \label{kf:eq1b}
\end{equation}
and the second term
\begin{eqnarray}
\Phi_{\text{I}} \left(12\right)&=& \phi_{\text{SW}} \left(r_{12}\right) \Psi\left(\hat{\mathbf{n}}_1,
\hat{\mathbf{n}}_2,\hat{\mathbf{r}}_{12} \right)
\label{kf:eq1c}
\end{eqnarray}
is the orientation-dependent attractive part which can be factorized into an isotropic square-well tail
\begin{equation}
\phi_{\text{SW}}\left(r\right)= \left\{ 
\begin{array}{ccc}
- \epsilon, &  &   \sigma<r< \lambda \sigma   \\ 
0,          &  &   \lambda \sigma < r \ %
\end{array}%
\right.\,\mbox{,} \label{kf:eq2}
\end{equation}
multiplied by an angular dependent factor
\begin{equation}
\Psi\left(\hat{\mathbf{n}}_1,
\hat{\mathbf{n}}_2,\hat{\mathbf{r}}_{12}\right)= \left\{ 
\begin{array}{ccccc}
1,    & \text{if}  &   \hat{\mathbf{n}}_1 \cdot \hat{\mathbf{r}}_{12} \ge \cos \theta_0 & \text{and} &  
-\hat{\mathbf{n}}_2 \cdot \hat{\mathbf{r}}_{12} \ge \cos \theta_0 \\ 
0,    &  & &\text{otherwise} & \
\end{array}%
\right.\,\mbox{.} \label{kf:eq3}
\end{equation}
The unit vectors $\hat{\mathbf{n}}_{i}(\omega_{i})$, ($i=1,2$), are defined by the 
spherical angles $\omega_i=(\theta_i,\varphi_i)$ in an arbitrarily oriented coordinate frame and $\hat{\mathbf{r}}_{12}(\Omega)$  is identified by the spherical 
angle $\Omega$ in the same frame. Reduced units, for temperature $T^*=k_B T/\epsilon$, pressure $P^{*}=\beta P/\rho$ and density $\rho^*=\rho \sigma^3$, 
will be used throughout, with $k_B$ being the Boltzmann constant. For future reference, we also introduce the packing fraction
$\eta=\pi \rho^{*}/6$. Two particles then attract if they are within the range of the square-well potential and if their attractive surfaces are properly aligned with each other, and repel as hard spheres otherwise. 

The relative ratio between attractive and total  surfaces 
is the  coverage $\chi$ that is related to the semi-angular width $\theta_0$ of the patch. This can be obtained as
\begin{eqnarray}
\label{kf:eq4}
\chi^2 &=& \left\langle \Psi\left(\hat{\mathbf{n}}_1,\hat{\mathbf{n}}_2,\hat{\mathbf{r}}_{12} 
\right) \right \rangle_{\omega_1 \omega_2} = \left\langle \Psi^{2}\left(\hat{\mathbf{n}}_1,\hat{\mathbf{n}}_2,\hat{\mathbf{r}}_{12} 
\right) \right \rangle_{\omega_1 \omega_2} =
\sin^4\left(\frac{\theta_0}{2}\right)\,\mbox{,}
\end{eqnarray}
\noindent where we have introduced the angular average
\begin{eqnarray}
\label{kf:eq5}
\left \langle \ldots \right \rangle_{\omega} &=& \frac{1}{4 \pi} \int d \omega \ldots\,\mbox{.}
\end{eqnarray}
\section{Berker-Henderson perturbation theory}
\label{sec:perturbation}
The Kern-Frenkel potential defined in Eqs. (\ref{kf:eq1}), (\ref{kf:eq1b}), (\ref{kf:eq1c}), (\ref{kf:eq2}), and (\ref{kf:eq3}),
leads to the natural separation into a reference one (the hard-sphere contribution) and an interaction term (the remaining, angular dependent,
part) that is usually requested by the standard perturbation theory prescription \cite{Gray84,Hansen86}. 

The presence of the hard-sphere potential for the reference part further suggests the Barker-Henderson (BH) scheme \cite{Barker67} 
as the most suitable
one for the present model. This has also the additional advantage that the free energy $F_0$ for the reference system can then be computed 
in several ways, as further discussed below.

The original scheme, due to Zwanzig \cite{Zwanzig54}, provided the first and second-order terms within the canonical ensemble, in the form
of a high temperature expansion
\begin{eqnarray}
\label{perturbation:eq4}
\frac{\beta \left(F - F_0\right)}{N} &=& \frac{\beta F_1}{N}+\frac{\beta F_2}{N}+ \ldots\,\mbox{,}
\end{eqnarray}
where the first term is proportional to $1/T^{*}$, the second to $(1/T^{*})^2$.

Although formally correct, it was noticed by Barker and Henderson \cite{Barker67,Henderson71,Barker76} that the corresponding expressions
were not useful for finite systems and a grand canonical ensemble derivation provided a much more convenient framework, where the
results for the canonical ensemble could be eventually obtained by a Legendre transformation.  

To the best of our knowldege, the details of the computation for the second-order term
were presented in Ref. \onlinecite{Henderson71}  only for isotropic potentials. As its generalization
to angular dependent potentials proves to be instructive, we have outlined
in Appendix \ref{sec:appa}.

The first term poses no problem and is computed in Eq.(\ref{appa:eq17}). When the perturbation parameter $\gamma=1$ and particularized to the Kern-Frenkel potential
given in Eqs. (\ref{kf:eq1}), (\ref{kf:eq1b}), (\ref{kf:eq1c}), (\ref{kf:eq2}), and (\ref{kf:eq3}) it becomes
\begin{eqnarray}
\label{perturbation:eq5}
\frac{\beta F_1}{N} &=& \frac{12 \eta}{\sigma^3} \int_{\sigma}^{\lambda \sigma}\! dr\, r^{2}  g_0\left(r\right)  \phi_{\text{SW}} \left(r\right) \left \langle
\beta \Psi\left(12\right) \right \rangle_{\omega_1,\omega_2}\,\mbox{.}
\end{eqnarray}
Note that this term is negative because so is $\phi_{\text{SW}}(r)$.

The second term is much more involved, but one can apply the same procedure as the isotropic case \cite{Goegelein08}, as detailed
in Appendix \ref{sec:appa}. The result for the second term is reported in Eq.(\ref{appa:eq18}). As in the isotropic case, however, this
derivation is of little practical use in view of the presence of the three and four point distribution functions \cite{Henderson71}.
Barker and Handerson \cite{Barker67}, devised then a simpler procedure to compute this term, based on a discrete representation
of the radial axis distributions. Again, the original procedure for spherically symmetric potentials is extended to angular dependent
potentials in Appendix \ref{sec:appb}.

The result for the second-order term is reported in Eq.(\ref{appb:eq15}). In case of the Kern-Frenkel potential, it yields
\begin{eqnarray}
\label{perturbation:eq9}
\frac{\beta F_2}{N} &=& -\frac{6 \eta}{\sigma^3}  \left(\frac{\partial \eta}{\partial P_0^{*}} \right)_T \int_{\sigma}^{\lambda \sigma}\! dr\, r^{2} 
g_0\left(r\right) \phi_{\text{SW}}^2 \left(r\right)
\left \langle \left[\beta \Psi\left(12\right) \right]^2 \right \rangle_{\omega_1,\omega_2}\,\mbox{,}
\end{eqnarray}
\noindent where $P_0^*=\beta P_0/\rho$ is the reduced pressure of the HS reference system in the Carnahan-Starling approximation
\cite{Carnahan69}.

This result is identical to that 
reported in Ref. \cite{Goegelein08} for a different radial part and it is known as \textit{macroscopic 
compressibility approximation} \cite{Barker67}.
Although the results given in Eqs.(\ref{perturbation:eq5}) (first order) and (\ref{perturbation:eq9}) (second order) 
are somewhat intuitive, being the natural extensions of the isotropic counterpart, a detail analysis of their derivations is
important as it might help to improve a drawback of the method that will be discussed at the end of Section \ref{sec:coexistence}.

\section{Fluid-Fluid and Fluid-Solid coexistence curves}
\label{sec:coexistence}
Once  the  reduced free energy per particle $\beta F/N$ is known, all thermodynamic properties can be derived. In particular, the pressures
and the chemical potentials can be derived from the standard thermodynamic identities \cite{Hansen86}
\begin{eqnarray}
\label{coexistence:eq1a}
\frac{\beta P}{\rho} &=& \eta \frac{\partial}{\partial \eta} \left(\frac{\beta F}{N} \right) \\
\label{coexistence:eq1b}
\beta \mu &=& \frac{\partial}{\partial \eta} \left(\eta \frac{\beta F}{N} \right)\,\mbox{.}
\end{eqnarray}

The gas-liquid (fluid-fluid) and fluid-solid coexistence curves are determined by the equality of the temperature, pressure and chemical potential 
in the two coexisting phases. 
Since the two phases are in contact, the condition on the equality of the temperature is always fulfilled. 
Thus, at fixed temperature $T^{*}$, we are left with the two conditions on the pressure and chemical potential.

For the gas-liquid coexistence, the conditions are
\begin{eqnarray}
\label{coexistence:eq2a}
P_{g}^{*} \left(T^{*},\rho^{*}_{g} \right)&=& P_{l}^{*} \left(T^{*},\rho^{*}_{l} \right) \\
\label{coexistence:eq2b}
\mu_{g}^{*} \left(T^{*},\rho^{*}_{g} \right)&=& \mu_{l}^{*} \left(T^{*},\rho^{*}_{l} \right)\,\mbox{,}
\end{eqnarray} 
where subscripts $g,l$ indicate that the quantity is computed in the gas and liquid phase respectively. The solution of this
system of non-linear equation gives $\rho_g^{*}=\rho_g^{*}(T^{*})$ for the gas coexistence branch, and
$\rho_l^{*}=\rho_l^{*}(T^{*})$ for the liquid coexistence branch. The hard-sphere reference part of the free energy 
(in excess with respect to the ideal gas) is assumed to be described by the Carnahan-Starling relation \cite{Carnahan69}
\begin{eqnarray}
\label{coexistence:eq3}
\left[\frac{\beta F_0}{N}\right]_{\text{liquid}} &=& \frac{4 \eta^2-3 \eta^3}{\eta \left(1-\eta\right)^2}\,\mbox{.}
\end{eqnarray}
For the hard-sphere radial distribution function $g_0(r)$ part appearing in Eqs. (\ref{perturbation:eq5}) and (\ref{perturbation:eq9}) the
Verlet-Weis \cite{Verlet72,Nagele04} corrected Percus-Yevick solution \cite{Wertheim63,Throop65} is exploited.
The details of the numerical procedure are reported in Appendix \ref{sec:appc}.

A similar method can be applied to the fluid-solid transition, resulting into the conditions
\begin{eqnarray}
\label{coexistence:eq4a}
P_{f}^{*} \left(T^{*},\rho^{*}_{f} \right)&=& P_{s}^{*} \left(T^{*},\rho^{*}_{s} \right) \\
\label{coexistence:eq4b}
\mu_{f}^{*} \left(T^{*},\rho^{*}_{f} \right)&=& \mu_{s}^{*} \left(T^{*},\rho^{*}_{s} \right)\,\mbox{.}
\end{eqnarray} 

All  calculations assume that the solid phase retains the crystal
structure of  the reference system, namely the face-centered
cubic (fcc) lattice. It is possible, especially at low $T$ or low $\chi$ where
anisotropy effects are more relevant, that fcc is not the most stable solid
for the model; our coexistence results are still valid, although possibly
relating to a metastable solid phase. We used   Wood's equation \cite{Wood52}
\begin{eqnarray}
\label{coexistence:eq5}
\left[\frac{\beta F_0}{N}\right]_{\text{solid}} &=& 2.1306  + 3 \ln \left( \frac{\eta}{1-\eta/\eta_{\text{cp}}}\right)+
\ln \left(\frac{\rho \Lambda^3}{\eta} \right)
\end{eqnarray}
\noindent for the solid free energy of the reference hard-sphere part, where $\eta_{cp}=\pi\sqrt{2}/6$ is the fcc volume fraction for closed packing.
For $g_{0}(r)$ in the solid phase, we use the orientation-averaged pair distribution function of Kincaid and Weis \cite{Kincaid77}.   

As a double check of the reliability of the numerical solution of Eqs.(\ref{coexistence:eq4a}) and (\ref{coexistence:eq4b}), 
the critical points were also computed using the alternative, and more direct method, as a maxima of the Helmholtz free energy, that is from the system
\begin{eqnarray}
\label{coexistence:eq6}
\frac{\partial^2}{\partial {{\rho^{*}}^2}} \left(\frac{\beta F}{N} \right) &=&0 \\ \nonumber
\frac{\partial^3}{\partial {{\rho^{*}}^3}} \left(\frac{\beta F}{N} \right) &=&0\,\mbox{.} \\
\end{eqnarray}
The solution provides $T^{*}_c$ and $\rho^{*}_c$ and are consistent with previous results, as they lay exactly on the top of the coexistence curves.
\section{Monte Carlo simulations}
\label{sec:montecarlo}
Standard Monte Carlo (MC) simulations in the NPT and in the 
grand-canonical (GC) ensembles have been implemented to evaluate the equation of state 
and the density dependence of the chemical potential for both the Janus and the SW model.
Translational and rotational moves consist of random translation of $\pm 0.1 \sigma$ and
random rotation of $\pm 0.1$ rad of a randomly selected particle.
In the GC study, insertion and deletion moves have been attempted, in average, every 500 displacement moves.  In NPT simulations, $N=500$ particles were investigated. In GC simulations box sizes were selected in such a way that  the number of particles 
in the simulation box was would always larger than 500. 
Fluid-fcc coexistence lines were calculated via
Gibbs-Duhem integration~\cite{Kofke93}, starting from initial coexistence values
at $T^{*}=2$ established via direct coexistence methods~\cite{vega-review}. Since at
infinite temperature the KF model reduces to the hard sphere model, coexistence
pressures at $T^{*}=2$, a very high value for the KF model, were searched for in the
vicinity of the  known HS values. We refer the reader to
Refs.~\cite{vega-review, Rom10a} for the details of the procedures. We point out
that all  $NPT$ simulations of the fcc solid were carried out in a cubic box
to constrain the system to retain the fcc arrangement
also in cases where the preferred structure would be a different one, possibly
other lattices or a distorted fcc. This choice was made to properly compare 
simulation results with the perturbation theory that assumes the cubic fcc
arrangement of the reference SW system.

\section{Results}
\label{sec:results}
\subsection{Equation of state and chemical potential}
\label{sec:eos}
In order to assess the performance of perturbation theory, we first compare results for pressure and chemical
potential as derived from the BH scheme outlined in Sect.\ref{sec:perturbation}, with numerical simulations
\cite{Sciortino09,Sciortino10,Giacometti09b,Giacometti10}. These values were further compared with those derived 
in Ref. \cite{Giacometti09b,Giacometti10} from integral equation (IE) theory within the reference hypernetted chain (RHNC),
following the method devised by Lado \cite{Lado82a,Lado82b,Lado82}. In the square-well case, integral equation values
were taken from Ref.\cite{Giacometti09a}.

\begin{figure}[htbp]
\begin{center}
\includegraphics[width=10cm]{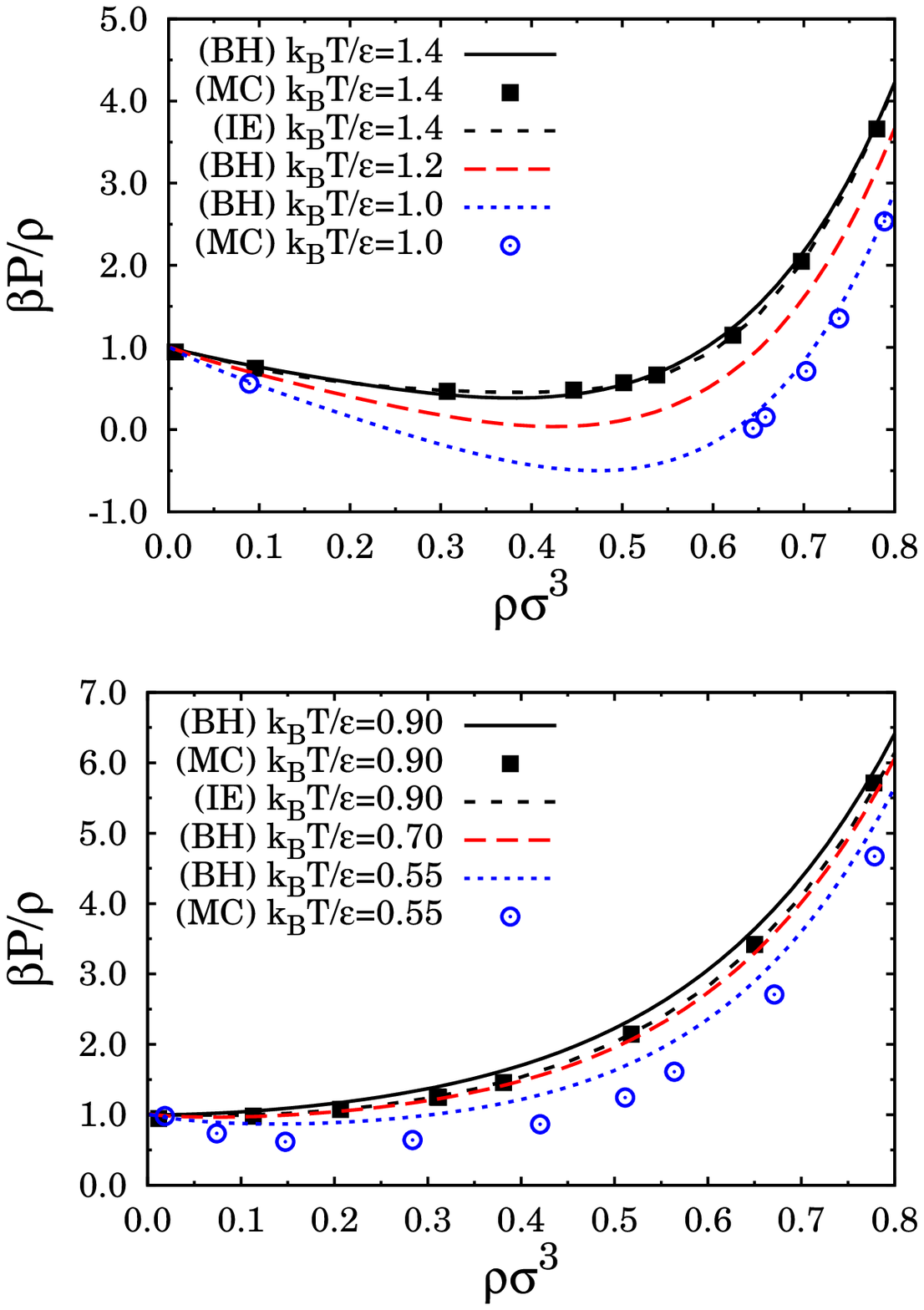} 
\end{center}
\caption{Reduced pressure $\beta P/\rho$ as a function of reduced density $\rho \sigma^3$
in the case of a square-well fluid with coverage $\chi=1$ (top panel), and in the case of a Janus fluid with coverage 
$\chi=0.5$ (bottom panel). A value of $\lambda=1.5$ is used. 
Results from BH thermodynamic perturbation theory (BH) are compared with Monte Carlo simulation 
(MC) and with RHNC integral equation theory (IE). Different curves refer to different temperatures as shown.}
\label{fig:fig2}
\end{figure}

\begin{figure}[htbp]
\begin{center}
\includegraphics[width=10cm]{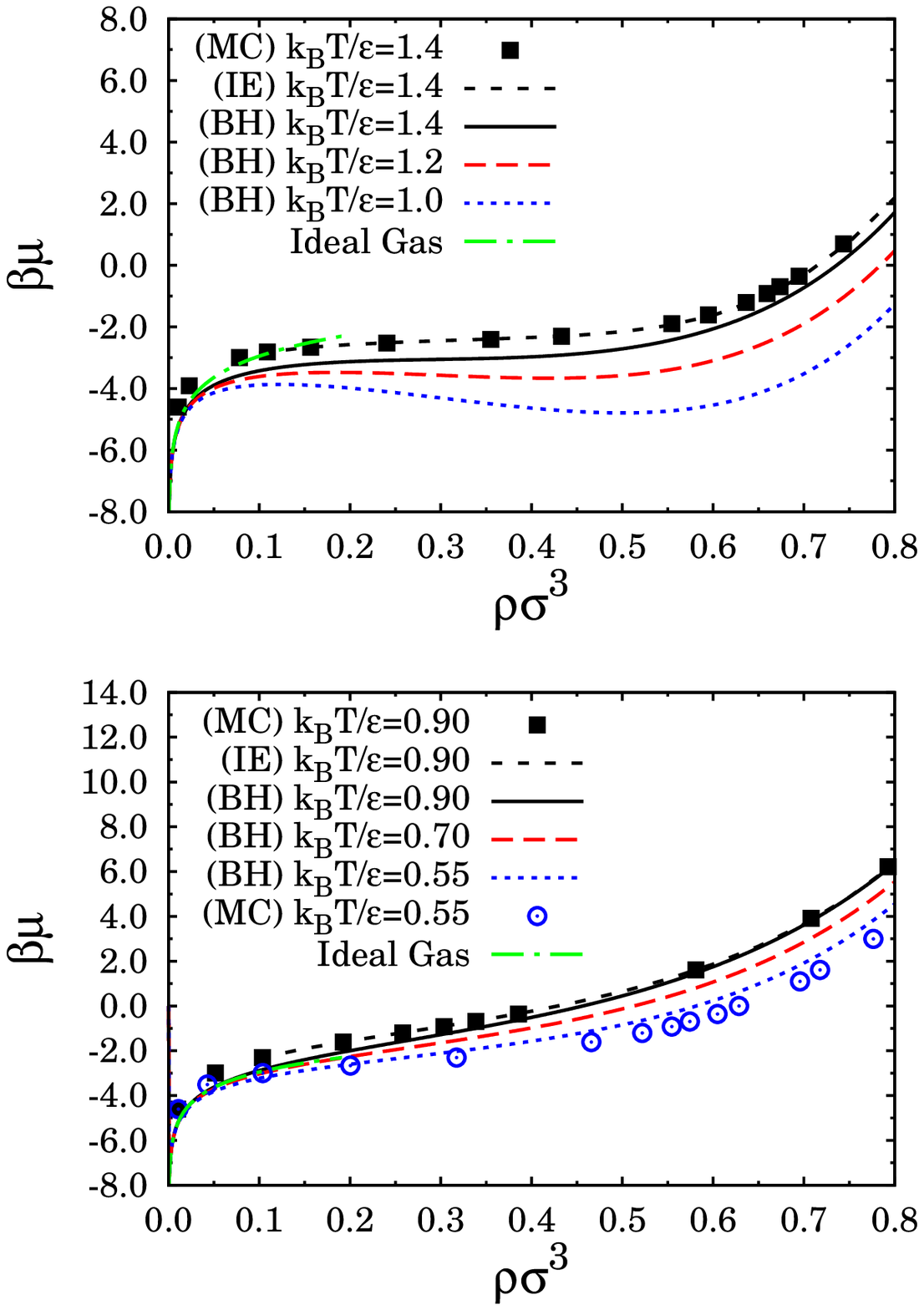}
\end{center}
\caption{Reduced chemical potential $\beta \mu $ as a function of reduced density $\rho \sigma^3$
in the case of a square-well fluid with coverage $\chi=1$ (top panel), and in the case of a Janus fluid with coverage 
$\chi=0.5$ (bottom panel).  A value of $\lambda=1.5$ is used. 
Results from BH thermodynamic perturbation theory (BH) are compared with Monte Carlo simulation 
(MC) and with RHNC integral equation theory (IE). Different curves refer to different temperatures as indicated. The low-density ideal gas limit
(light dashed line) is also depicted.}
\label{fig:fig3}
\end{figure}

The results are shown in Figures \ref{fig:fig2} (pressure) and \ref{fig:fig3} (chemical potential), 
for two representative values of the
coverages, namely the square-well ($\chi=1$) (top panels) and the Janus ($\chi=0.5$) (bottom panels). In all cases, a value of $\lambda =1.5$
for the total extension of the well (in units of the hard-spheres diameter), was selected in order to compare with
previous results.

In the square-well case (top panels), the first selected temperature $k_B T/\epsilon=1.4$ corresponds to a temperature above the
critical temperature, while the last one $k_B T/\epsilon=1.0$ is well below (see Ref.\cite{Giacometti09a} and references therein).
In both cases, the performance of the BH thermodynamic perturbation theory  is remarkably good, in agreement with
previous findings on the square-well potential \cite{Henderson80}. The dip in the curve for $k_B T/\epsilon=1.0$ indeed corresponds
to the van der Waals loop typical of the coexistence region. In the case of chemical potential (Fig.\ref{fig:fig3}), the ideal gas low density solution
$\beta \mu = \ln(\rho \sigma^3)$ is also reported for comparison. Interestingly, while both Monte Carlo (MC) simulations
and RHNC integral equation theory (IE) converge to the correct limit, the BH perturbation theory appears to underestimate the chemical
potential in the whole range of densities. On the other hand, it provides the same quality results for all temperatures, even in those regions
where integral equation theory are known to experience difficulties.

Slightly less satisfactory results are obtained in the case of a Janus fluid, as shown in the bottom panels of both 
Figs. \ref{fig:fig2} and \ref{fig:fig3}.
Here the two limiting temperatures $k_B T/\epsilon=0.9$ and $k_B T/\epsilon=0.55$ are both above the critical temperature \cite{Sciortino10}, as
apparent from the absence of any loop in both the pressure and the chemical potential. The Janus phase diagram, however, is known to be anomalous
\cite{Sciortino09}, as a result of a competition with a micelle formation process that destabilizes the condensation one \cite{Sciortino10}.
In this case the BH thermodynamic perturbation theory (BH) does not show a well defined pattern as it overestimates the pressure
for both temperatures (Fig.\ref{fig:fig2} bottom panel), 
as well as the chemical potential for $k_B T/\epsilon=0.55$, but underestimates it for  the higher temperature 
$k_B T/\epsilon=0.9$ (Fig. \ref{fig:fig3} bottom panel). While it is known that the BH compressibility approximation
can be expected to display different performance at different densities due the presence of higher-order terms   \cite{Barker67},
the above behavior is more likely to be attributed to the anomalous behavior of the Janus phase diagram that perturbation theory cannot
capture at the present stage. In spite of this, the performance of BH thermodynamic perturbation theory remains remarkable, 
especially in view of the difficulties experienced by
integral equation theories at such low temperatures associated with low surface coverages.
\subsection{The fluid-fluid coexistence}
\label{sec:ff}
\begin{figure}[htbp]
\begin{center}
\includegraphics[width=12cm]{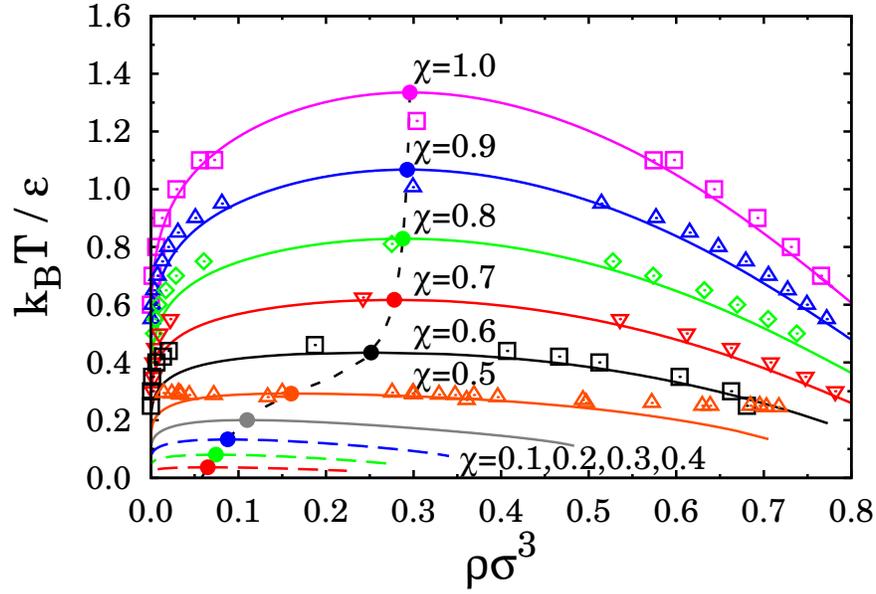}
\end{center}
\caption{The fluid-fluid coexistence curves as computed from BH perturbation theory
and compared against numerical simulations. Lines are from perturbation theory, points from numerical simulations, for $\lambda=1.5$ 
from Ref.~\cite{Sciortino10}. 
All coverages from $\chi=1.0$ (SW case) to $\chi=0.0$ (HS case) are depicted
in the former case, whereas numerical simulations are in the range $0.5 \le \chi \le 1.0$, that is from the Janus
to the SW limit. }
\label{fig:fig4}
\end{figure}

A very stringent test on the reliability of BH thermodynamic perturbation theory stems from the calculation of the fluid-fluid (gas-liquid) coexistence
curves. This is depicted in Figure \ref{fig:fig4} where the coexistence curves are computed by BH thermodynamic perturbation theory
(solid lines), and contrasted with results from Monte Carlo numerical simulations  (points), 
from Ref.~\cite{Sciortino10}. The considered coverages range from $\chi=1.0$, corresponding
to the SW fluid, to $\chi=0.5$, corresponding to the Janus limit, and are identical to those considered in Ref.\cite{Sciortino10}.
As before, $\lambda=1.5$ was assumed in all calculations. In the BH thermodynamic perturbation theory, further coverages down to the hard-sphere limit
were also computed. In all cases, the critical points stemming from the BH calculations are also displayed as solid circles on the binodals.

The performance of the BH method appears to be remarkably good.  Both the vapor and the liquid branches of the numerical simulations are closely followed
by the BH calculations, with an accuracy almost independent of the considered coverage, with the only exception of the Janus case ($\chi=0.5$) that
is however known to have an anomalous behavior \cite{Sciortino09}, as remarked. This is only apparently in contrast with results from chemical potential, 
reported in previous Section \ref{sec:eos}, where the BH results for chemical potential in the Janus case appeared to be less precise than in the SW case.
On the one hand, a closer inspection reveals that BH results for each single coverage do indeed show a small quantitative discrepancy with the corresponding 
MC simulation, more or less uniform in the entire density-temperature plane. On the other hand, this latter feature constitutes an advantage in the method
as a numerical solution of Eq.(\ref{coexistence:eq2b}) may provide accurate coexistence lines if both the vapor and the liquid chemical potentials
have similar inaccuracies. This results is, nonetheless, comparable in accuracy with those stemming from reference hypernetted chain (RHNC) integral
equation theory \cite{Giacometti09a,Giacometti09b,Giacometti10}, with the additional advantage of a less computational and algorithmical complexity
and, more importantly, of being able to access the critical region, including the critical point, that is one of the main shortcomings common to all integral
equation theories.

It is worth noticing how BH perturbation theory can provide an accurate prediction of the location of the coexistence lines even below the Janus limit,
that is for $\chi<0.5$, where extensive numerical simulations are so-far suggesting the fluid-fluid transition to be inhibited by
a micellization process \cite{Sciortino10}. This could be useful for a future more focussed numerical calculation within a limited region of the phase
diagram where   BH theory predicts coexistence to occur. 
\subsection{The fluid-solid coexistence}
\label{sec:fs}
Let us now turn to the fluid-solid coexistence, a calculation that has not been carried out so far for this model by any method.
As illustrated in Sec.\ref{sec:coexistence} and below, BH perturbation scheme allows this analysis with an effort, both computational and algorithmical, 
comparable with that of the fluid-fluid case.

In the isotropic SW case with $\lambda=1.5$, the reference point for this calculation are those obtained as early as in 1980 by Young and Adler \cite{Young80}. 
Using molecular dynamics (MD), they reported a detailed study of the different crystal structures (fcc, hcp, and bcc) with the corresponding Helmholtz free energies,
thus arguing that fcc and hcp were the most stable structures within the entire temperature-density plane.
Additional numerical simulations were later performed by Marr and Gast \cite{Marr93}, Serrano-Ill\'an and Navascu\'es \cite{Serrano06}, and Kiselev {\it et al.} \cite{Kiselev06} essentially confirming this scenario. A very detailed study of the entire phase diagram of the SW, was carried out by Liu {\it et al.}
\cite{Liu05}.

\begin{figure}[htbp]
\begin{center}
\includegraphics[width=12cm]{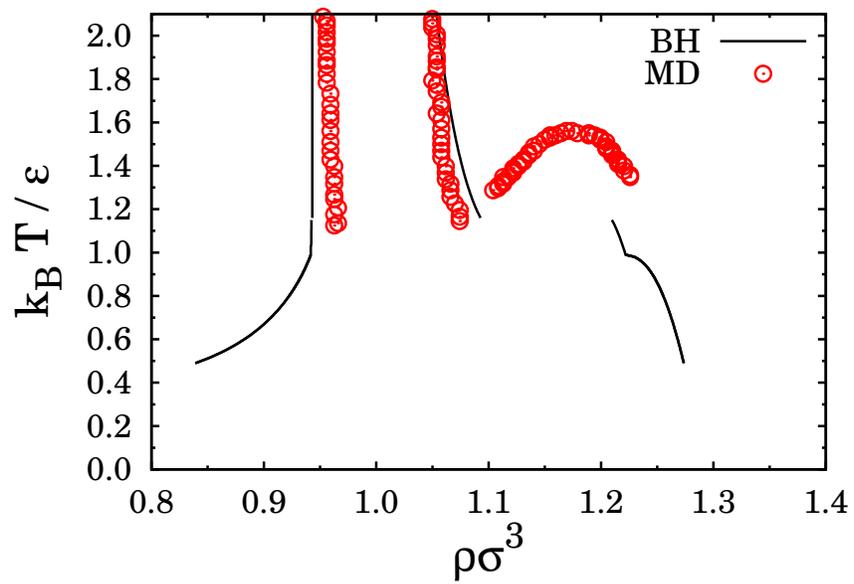}
\end{center}
\caption{Fluid-solid coexistence for the case of the SW potential ($\chi=1.0$) with $\lambda=1.5$. Results from Barker-Henderson (BH) perturbation
theory are contrasted with molecular dynamics (MD) data by Young and Adler \cite{Young80}.}
\label{fig:fig5}
\end{figure}

In Fig.\ref{fig:fig5}, we report results from  BH   thermodynamic perturbation theory (solid line) along with results
from Young and Adler (circles).

While at high temperatures all calculations agree, discrepancies start to appear on cooling away from the hard spheres limit.
In particular, the plateau appearing in the solid branch of MD calculations indicates a fcc-fcc (or fcc-hcp) transition that is not
accounted for in BH calculations, that assumed fcc structures all the way, although in principle it could be done.  
In the BH calculations, in particular, the difficulty arises from the stability of the numerical scheme used for the solution
of Eqs.(\ref{coexistence:eq4a}) and (\ref{coexistence:eq4b}).

\begin{figure}[htbp]
\begin{center}
\includegraphics[width=12cm]{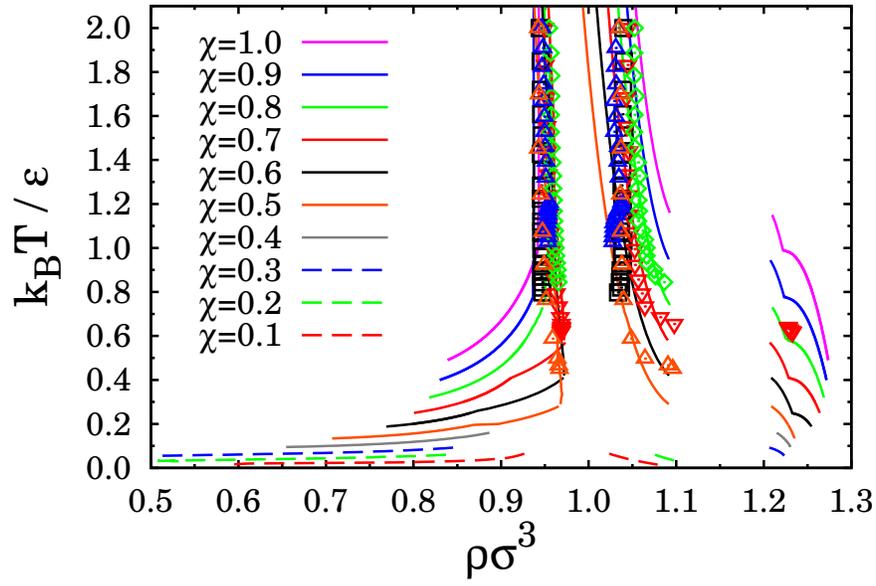}
\end{center}
\caption{Coverage dependence of the fluid-solid coexistence curves. Again $\lambda=1.5$ and considered coverages are from
$\chi=1.0$ (SW case) to $\chi=0.1$ for Barker-Henderson perturbation theory (lines) and from $\chi=0.9$ to $\chi=0.5$ (Janus)
for Monte Carlo simulations.}
\label{fig:fig6}
\end{figure}

For lower coverages, no previous calculations on the Kern-Frenkel model exist to compare with. Fig.\ref{fig:fig6} illustrate the coverage dependence
of the fluid-solid coexistence lines as computed from MC simulations (points) and from BH thermodynamic perturbation theory (lines). 
As in the fluid-fluid case, 
MC simulations have been obtained up to the Janus fluid ($\chi=0.5$), whereas BH theory provides results even below that limit. 
Simulations below the Janus limit could be done, but are computationally more demanding.

As in the SW case, even for lower coverages one might expect a structural transition at a certain density. Even assuming fcc to be the most 
stable structure,
the range of the potential associated with the value $\lambda=1.5$ used here, allows a fcc-fcc transition between one fcc with only 
nearest-neighbors
bonded, and a more denser one with even the next-to-nearest-neighbors are bonded. This is associated with the jump in density that is
present in some of the plots of Fig.\ref{fig:fig6}.

\section{Conclusions}
\label{sec:conclusions}
In this paper we presented the first Barker-Henderson (BH) perturbative calculation for the one-patch Kern-Frenkel model, 
and compared with specialized MC simulations.
The BH method hinges on a second-order thermodynamic perturbation theory in the inverse temperature, allowing the calculation of the Helmholtz 
free energy within this approximation, and hence,
of all thermodynamic quantities, such as the density and temperature dependence of pressure and chemical potential.
A numerical solution has then been implemented to infer the fluid-fluid coexistence line (binodal) from the equality of pressure and chemical potential 
in the vapor and liquid phase at a given temperature. A similar procedure also provides the fluid-solid transition.

When compared with numerical simulations, the BH predictions are found to be extremely reliable in the entire phase diagram, and for all coverages
from the isotropic SW potential to the lowest considered coverage ($\chi=0.1$) very close to the hard-spheres limit. This constitutes  one of the
main advantages with respect to, in principle, superior and more accurate theoretical methodologies hinging on integral equation solutions, that
are typically affected by the impossibility of accessing the critical region, and by the numerical instabilities occurring at low temperatures
associated with low coverages. Even at the quantitative level, BH results were found to be competitive with integral equation theories, 
in agreement with previous results on the isotropic SW fluid.

The performance of BH is particularly noteworthy for coverages below the Janus limit, that is for $\chi<0.5$, the most challenging region for
numerical simulations in view of the tendency for the particles to form single and multi-layer clusters always exposing the hard-sphere
surface in the outer region in order to maximize favorable contacts. This mechanism competes and destabilizes the condensation process and
the interpretation of numerical simulation results become more and more obscure in that region. As a result, a clear scenario suggested
by numerical simulations in this region is still missing. A better understanding could in principle be favored by our BH results that 
provide a well defined and restricted
region of the temperature-density plane where indication of possible coexistences could be sought for.

While in the present paper the BH method has been applied to a single patch Kern-Frenkel potential, the method could potentially be extended to
higher number of patches with no difficulties. As a matter of fact, this has already been done in Ref.\cite{Goegelein08} for two-patch colloids with
Yukawa interactions for the attractive part. An  inspection of the relevant equations (\ref{perturbation:eq5}) and (\ref{perturbation:eq9}), however,
suggest the result to be identical to the one-patch case at the same coverage. This means that the BH method, in the present form, is not
capable of distinguishing between one and two-patches, at the same coverage, a feature that, conversely, is accounted for in both numerical simulations
\cite{Sciortino09,Sciortino10} and integral equation theory \cite{Giacometti09b,Giacometti10}. In particular, it cannot then account for the anomalous
behavior present in the Janus limit of the single patch \cite{Sciortino09} and not present in the corresponding coverage ($\chi=0.5$) of the
double patches model \cite{Sciortino10}. This is rather surprising in view of the fact that a similar method,
based on a low-density virial expansion, applied to a companion problem, was able to distinguish between single and double patches, albeit with
a rather poor estimate for the fluid-fluid transition \cite{Fantoni07}. A promising approach in this respect appears to be
the perturbative scheme devised for molecular fluids by Gubbins, Gray and others \cite{Gubbins72,Gray84}, who considered
an expansion in powers of the anisotropic part of the potential, in a way akin to that discussed in Appendix \ref{sec:appa}, often supplemented by
a Pad\'e approximant to improve the convergence of the expansion, as proposed by Stell et al \cite{Stell74}.

We plan to investigate this and other points in details in future work.

\begin{acknowledgments}
CG acknowledges the hospitality of Universit\`a Ca' Foscari of Venice where this work was started.  
FS and FR acknowledge support from ERC-226207-PATCHYCOLLOIDS. 
\end{acknowledgments}
\appendix
\section{The second order perturbation theory} 
\label{sec:appa}
As explained in Ref.\cite{Henderson71}, the most correct way of developing a perturbation expansion  is in the grand-canonical ensemble.
Assume a general potential of the form
\begin{eqnarray}
\label{appa:eq1}
U_{\gamma}\left(1,\ldots,N\right)&=& U_{0}\left(1,\ldots,N\right)+\gamma U_{I}\left(1,\ldots,N\right)\\ \nonumber 
&=&\sum_{i<j} \Phi_{\gamma}\left(ij\right) = \sum_{i<j} \Phi_{0}\left(ij\right)+\gamma \sum_{i<j} \Phi_{\text{I}}\left(ij\right)\,\mbox{,}
\end{eqnarray}
where $U_{0}(1,\ldots , N)=\sum_{i,j} \Phi_{0}(ij)$ is the unperturbed part and $U_{I}(1,\ldots ,N)=\sum_{i,j} \Phi_{\text{I}}(ij)$
is the perturbation part. Here $0\le \gamma \le 1$ is used as perturbativ parameter, and each coordinate $i$ includes both
the coordinate $\mathbf{r}_i$ and patch orientation $\hat{\mathbf{n}}_i$, so that  $i\equiv (\mathbf{r}_i,\hat{\mathbf{n}}_i).$
Also, $\beta=1/(k_B T)$ denotes the inverse of the thermal energy.

Introducing the following short-hand notation 
\begin{eqnarray}
\label{appa:eq1b}
\int_{1,\ldots,N} \left(\cdots \right) &\equiv & \int \left[\prod_{i=1}^N d \mathbf{r}_i \left \langle \left( \cdots \right) 
\right \rangle_{\omega_{i}} \right]
\end{eqnarray}
for the integration over all particle coordinates, the grand-canonical partition function 
\begin{eqnarray}
\label{appa:eq2}
{\cal Q}_{\gamma} &=& \sum_{N=0}^{+\infty} \frac{e^{\beta \mu N}}{N!\Lambda_T^{3N} } \int_{1,\ldots,N} e^{-\beta U_{\gamma}}=
e^{-\beta \Omega_{\gamma}}
\end{eqnarray}
(here $\Lambda_T$ is the de Broglie thermal wavelength, and $\Omega_{\gamma}$ is the grand-potential) 
can then be used to obtain an expansion of the Helmholtz free energy
\begin{eqnarray}
\label{appa:eq3}
F_{\gamma} &=& F_0+ \gamma \left(\frac{\partial F_{\gamma}}{\partial \gamma} \right)_{\gamma=0}+
\frac{1}{2!} \gamma^2 \left(\frac{\partial^2 F_{\gamma}}{\partial \gamma^2} \right)_{\gamma=0}+ \cdots
\end{eqnarray}
as follows \cite{Henderson71}.

Taking the derivative of $\ln {\cal Q}_{\gamma}$ at fixed chemical potential $\mu$, one has, using Eq.(\ref{appa:eq1})
\begin{eqnarray}
\label{appa:eq4}
\left[ \frac{\partial}{\partial \gamma} \ln {\cal Q}_{\gamma} \right]_{\mu} &=& \frac{1}{2} \int_{1,2} \frac{\partial}{\partial \gamma}
\left[-\beta \Phi_{\gamma}  \left(12\right) \right] \rho_{\gamma} \left(12\right)\,\mbox{,}
\end{eqnarray}
where
\begin{eqnarray}
\label{appa:eq5}
\rho_{\gamma} \left(1\ldots h\right)&=& \frac{1}{{\cal Q}_{\gamma}}  
\sum_{N=h}^{+\infty} \frac{e^{\beta \mu N}}{\left(N-h\right)!\Lambda_T^{3N} } \int_{1,\ldots,N} e^{-\beta U_{\gamma}}\,\mbox{.}
\end{eqnarray}
The second derivative is somewhat more laborious \cite{Henderson71}, and one obtains after some algebra
\begin{eqnarray}
\label{appa:eq6}
\left[ \frac{\partial^2}{\partial \gamma^2} \ln {\cal Q}_{\gamma} \right]_{\mu} &=& \frac{1}{2} \int_{1,2} \frac{\partial^2}{\partial \gamma^2}
\left[ -\beta \Phi_{\gamma} \left(12\right) \right] \rho_{\gamma} \left(12\right)+\frac{1}{2} \int_{1,2} 
\left(\frac{\partial}{\partial \gamma}\left[ -\beta \Phi_{\gamma} \left(12\right) \right]\right)^2 \rho_{\gamma} \left(12\right) \\ \nonumber
&+& \int_{1,2,3}  \frac{\partial}{\partial \gamma}
\left[-\beta \Phi_{\gamma}  \left(12\right) \right]  \frac{\partial}{\partial \gamma}
\left[-\beta \Phi_{\gamma}  \left(23\right) \right] \rho_{\gamma} \left(123\right) \\ \nonumber
&+& \frac{1}{4} \int_{1,2,3,4} \frac{\partial}{\partial \gamma} \left[-\beta \Phi_{\gamma}  \left(12\right) \right]  
\frac{\partial}{\partial \gamma}
\left[-\beta \Phi_{\gamma}  \left(34\right) \right] \left[\rho_{\gamma}\left(1234\right)-\rho_{\gamma} \left(12\right)  \rho_{\gamma} 
\left(34\right) \right]\,\mbox{.}
\end{eqnarray}

The free energy $F_{\gamma}$ is then obtained by considering $\gamma$ as an additional thermodynamical variable, and by
performing the appropriate manipulations \cite{Henderson71}. One then has
\begin{eqnarray}
\label{appa:eq7}
F_{\gamma} &=& \mu N - k_B T \ln {\cal Q}_{\gamma}\,\mbox{,}
\end{eqnarray}
and
\begin{eqnarray}
\label{appa:eq8}
N&=& k_B T \left[ \frac{\partial}{\partial \mu} \ln {\cal Q}_{\gamma} \right]_{\gamma}\,\mbox{,}
\end{eqnarray}
where, for notational simplicity, here we do not distinguish between the canonical and grand-canonical
number of particles $N$. Then
\begin{eqnarray}
\label{appa:eq9}
- k_B T \left[\frac{\partial}{\partial \gamma} \ln {\cal Q}_{\gamma} \right]_{\rho} &=& 
\left(\frac{\partial \Omega_{\gamma}}{\partial \gamma}\right)_{\rho} = 
\left(\frac{\partial \Omega_{\gamma}}{\partial \gamma}\right)_{\mu}-
\left(\frac{\partial \Omega_{\gamma}}{\partial \mu}\right)_{\gamma}
\left(\frac{\partial \rho}{\partial \gamma} \right)_{\mu}
\left(\frac{\partial \mu}{\partial \rho} \right)_{\gamma}
\end{eqnarray}
and hence, using the chain rule
\begin{eqnarray}
\label{appa:eq10}
\left(\frac{\partial \rho}{\partial \gamma} \right)_{\mu} 
\left(\frac{\partial \gamma}{\partial \mu} \right)_{\rho}
\left(\frac{\partial \mu}{\partial \rho} \right)_{\gamma} &=& -1,
\end{eqnarray}
one gets
\begin{eqnarray}
\label{appa:eq11}
-k_B T \left[\frac{\partial}{\partial \gamma} \ln {\cal Q}_{\gamma} \right]_{\rho} &=&
-k_B T \left[\frac{\partial}{\partial \gamma} \ln {\cal Q}_{\gamma} \right]_{\mu}
-k_B T \left[\frac{\partial}{\partial \mu} \ln {\cal Q}_{\gamma} \right]_{\gamma}
\left(\frac{\partial \mu}{\partial \gamma} \right)_{\rho} 
\end{eqnarray}
that, with the help of Eq.(\ref{appa:eq8}), leads to
\begin{eqnarray}
\label{appa:eq12}
\left(\frac{\partial F_{\gamma}}{\partial \gamma} \right)_{\rho} &=&
-k_B T \left[ \frac{\partial}{\partial \gamma} \ln {\cal Q}_{\gamma} \right]_{\mu}\,\mbox{,}
\end{eqnarray}
where the right-hand-side is given by Eq.(\ref{appa:eq4}).

For the second derivative, one proceeds as before, to obtain
\begin{eqnarray}
\label{appa:eq13}
\left(\frac{\partial^2 F_{\gamma}}{\partial \gamma^2} \right)_{\rho} &=& -k_B T
\left(\frac{\partial^2}{\partial \gamma^2} \ln {\cal Q}_{\gamma} \right)_{\mu}+
k_B T \left[\left(\frac{\partial^2}{\partial \gamma \partial \mu} \ln {\cal Q}_{\gamma} \right) \right]_{\gamma}^2
/ \left[\left(\frac{\partial^2}{\partial \mu^2} \ln {\cal Q}_{\gamma} \right) \right]_{\gamma}\,\mbox{.}
\end{eqnarray}
Using Eq.(\ref{appa:eq8}) and the relation
\begin{eqnarray}
\label{appa:eq14}
\frac{\partial}{\partial \mu} &=& \rho \left(\frac{\partial \rho}{\partial P} \right) \frac{\partial}{\partial \rho}\,\mbox{,}
\end{eqnarray}
one finds
\begin{eqnarray}
\label{appa:eq15}
-k_B T \left[\frac{\partial^2}{\partial \gamma \partial \mu}  \ln {\cal Q}_{\gamma} \right] &=&
\rho \left(\frac{\partial \rho}{\partial P} \right) \frac{\partial}{\partial \rho} 
\left[\frac{1}{2} \frac{\partial}{\partial \gamma}
\left[-\beta \Phi_{\gamma}  \left(12\right) \right] \rho_{\gamma} \left(12\right)\right]\,\mbox{.}
\end{eqnarray}
Substituting in Eq.(\ref{appa:eq13}), one finds
\begin{eqnarray}
\label{appa:eq16}
\left(\frac{\partial^2 F_{\gamma}}{\partial \gamma^2} \right)_{\rho}&=& -k_B T
\left(\frac{\partial^2}{\partial \gamma^2} \ln {\cal Q}_{\gamma} \right)_{\mu}+
\frac{N}{V^2} \left(\frac{\partial \rho}{\partial P} \right)
\left\{
\frac{\partial}{\partial \rho} 
\left[\frac{1}{2} \frac{\partial}{\partial \gamma}
\left[-\beta \Phi_{\gamma}  \left(12\right) \right] \rho_{\gamma} \left(12\right)\right]
\right\}^2\,\mbox{,}
\end{eqnarray}
where the first term on the right-hand-side is given by Eq.(\ref{appa:eq6}).

The first and second order solutions, can be finally particularized to the potential form
given in Eq.(\ref{appa:eq1}), so that Eqs.(\ref{appa:eq4}) and (\ref{appa:eq12}) lead to
\begin{eqnarray}
\label{appa:eq17}
\left[\frac{\partial}{\partial \gamma} \left(\beta F_{\gamma} \right) \right]_{\gamma=0} &=&
\frac{1}{2} \rho N \int d \mathbf{r}_{12} \left \langle \beta \Phi_{\text{I}}\left(12\right) \right \rangle_{\omega_1,\omega_2}
g_0\left(12\right)
\end{eqnarray}
and Eqs.(\ref{appa:eq6}) and (\ref{appa:eq16}) leads to
\begin{eqnarray}
\label{appa:eq18}
 \left(\frac{\partial^2 }{\partial \gamma^2}\left(\beta F_{\gamma}\right) \right)_{\gamma=0}&=& -\frac{1}{2} N \rho 
\int d \mathbf{r}_{12}  \left \langle \left[ -\beta \Phi_{\text{I}} \left(12\right) \right]^2 \right \rangle_{\omega_1,\omega_2}
g_0\left(12\right) \\ \nonumber
&-& N \rho^2 \int d \mathbf{r}_{12} d\mathbf{r}_{13}    \left \langle
\left[-\beta \Phi_{\text{I}}  \left(12\right) \right]  
\left[-\beta \Phi_{\text{I}}  \left(23\right) \right] \right \rangle_{\omega_1,\omega_2,\omega_3} g_{0} \left(123\right) \\ \nonumber
&-& \frac{1}{4} N \rho^3 \int d \mathbf{r}_{12} d\mathbf{r}_{13} d\mathbf{r}_{14} \left \langle 
\left[-\beta \Phi_{\text{I}}  \left(12\right) \right]  
\left[-\beta \Phi_{\text{I}}  \left(34\right) \right] \right \rangle_{\omega_1,\omega_2,\omega_3,\omega_4}
\left[g_{0}\left(1234\right)-g_{0} \left(12\right)  g_{0} \left(34\right) \right] \\ \nonumber
&+& \beta N \left(\frac{\partial \rho}{\partial P} \right) 
\left\{
\frac{\partial}{\partial \rho} \left[ \frac{1}{2} \rho^2 \int d \mathbf{r}_{12} \left \langle \Phi_{\text{I}}\left(12\right)
\right \rangle_{\omega_1,\omega_2} g_{0}\left(12\right) \right]
\right\}^2\,\mbox{.}
\end{eqnarray}
\section{The Barker-Henderson discrete representation} 
\label{sec:appb}
As in the spherically potential case, the above expressions are, however, not very useful for practical computation, due to the
high complexity involved in the calculations of the three $g_{0}(123)$ and four $g_{0}(1234)$ point correlation functions.

Following the original work by Barker and Henderson, we return to the canonical partition function
\begin{eqnarray}
\label{appb:eq1}
Q &=& \frac{1}{N! \Lambda_T^{3N}} \int_{1,\ldots,N} e^{-\beta U\left(1,\ldots,N\right)}=\frac{1}{N! \Lambda_T^{3N}} Z=e^{-\beta F}
\end{eqnarray}
that is related to the configurational partition function $Z$ and to the Helmholtz free energy $F$. The intermolecular distance axis $r_{ij}$
is divided in intervals $(0,r_1),(r_1,r_2),\ldots,(r_l,r_{l+1}),\ldots$ in such a way that there are $N_{l}$ distances in the $l-$th interval
$(r_l,r_{l+1})$. The total potential $U$ appearing in {Eq.}(\ref{appb:eq1}) can then be written as a sum over the different intervals
with the respective multiplicity
\begin{eqnarray}
\label{appb:eq2}
U\left(1,\ldots,N\right)&=& \sum_{l} N_{l} \overline{\Phi}\left(r_{l},\left\{\Omega,\omega\right\}_{l}\right)\,\mbox{,}
\end{eqnarray} 
where $\overline{\Phi}(r_{l},\{\Omega,\omega\}_{l})$ is the average potential in the $l-$th interval (assumed to be constant),
and $\{\Omega,\omega\}_{l}$ are the set of orientational angles included in the same interval.

Again we assume that the potential can be split into a hard-sphere part plus a tail
\begin{eqnarray}
\overline{\Phi}\left(r_{l},\left\{\Omega,\omega\right\}_{l}\right)&=&\overline{\phi}_{0}\left(r_{l} \right)+
\overline{\Phi}_{\text{I}}\left(r_{l},\left\{\Omega,\omega\right\}_{l}\right)\,\mbox{.}
\label{appb:eq3}
\end{eqnarray}

Introducing 
the average over the unperturbed system having $Z_0$ as configurational partition function
\begin{eqnarray}
\label{appb:eq5}
\left \langle \ldots \right \rangle_{0} &=& \frac{1}{Z_{0}} \sum_{N_{1},N_{2},\ldots} \int_{R} d \mathbf{r}_{1} \cdots d \mathbf{r}_{N}
e^{-\beta \sum_{l} N_{l}  \overline{\phi}_{0}\left(r_{l} \right)}\,\mbox{,}
\end{eqnarray}
where the symbol $R$ indicates that the integral is restricted to configurations having $N_{l}$ intermolecular distances in the
interval $(r_{l},r_{l+1})$, the Helmholtz free energy can be written in terms of that of hard-spheres $F_0$ as
\begin{eqnarray}
\label{appb:eq6}
\beta F &=& \beta F_{0} - \ln \left \langle \left \langle e^{-\beta \sum_{l} N_{l}  \overline{\Phi}_{\text{I}}\left(r_{l},\left\{\Omega,
\omega \right\}_{l} \right)} \right \rangle_{\{\omega\}} \right \rangle_{0}\,\mbox{.}
\end{eqnarray} 
Note that the angular average over the $\{\Omega\}$ variables is included in the average (\ref{appb:eq5}).

Use of the cumulant expansion 
\begin{eqnarray}
\label{appb:eq7}
-\ln \left \langle e^{-\lambda x} \right \rangle &=& \lambda \left \langle x \right \rangle 
- \frac{1}{2} \lambda^2 \left( \left \langle x^2 \right \rangle - \left \langle x \right \rangle^2 \right)+ \ldots
\end{eqnarray}
leads to 
\begin{eqnarray}
\label{appb:eq8}
\beta \left(F-F_{0} \right) &=& \beta F_{1} + \beta F_{2} + \ldots\,\mbox{,}
\end{eqnarray}
where
\begin{eqnarray}
\label{appb:eq9}
\beta F_1 &=& \sum_{l} \left \langle \left \langle N_{l}   
\beta \overline{\Phi}_{\text{I}} \left(r_{l},\left\{\Omega,\omega \right\}_{l} \right)  
\right \rangle_{\{\omega\}} \right \rangle_{0}\,\mbox{,}
\end{eqnarray}
and where
\begin{eqnarray}
\label{appb:eq10}
\beta F_{2} &=& - \frac{1}{2} \sum_{lm} \left \langle  \left \langle N_{l} N_{m}  
\beta \overline{\Phi}_{\text{I}} \left(r_{l},\left\{\Omega,\omega \right\}_{l} \right) 
\beta \overline{\Phi}_{\text{I}} \left(r_{m},\left\{\Omega,\omega \right\}_{m} \right)
\right \rangle_{\{\omega\}} \right \rangle_{0}\,\mbox{.}
\end{eqnarray}
As \cite{Barker67},
\begin{eqnarray}
\label{appb:eq11}
\left \langle N_{l} \right \rangle_{0} &=& 2 \pi \rho N \int_{r_{l}}^{r_{l+1}} dr r^2 g_{0} \left(r\right)\,\mbox{,}
\end{eqnarray} 
the first order term becomes 
\begin{eqnarray}
\label{appb:eq12}
\beta F_{1} &=& \frac{1}{2} \rho N \int d \mathbf{r}g_{0} \left(r\right) 
\left \langle \beta \Phi_{\text{I}} \left(r,\Omega,\omega_{1},\omega_{2}  \right)
\right \rangle_{\omega_{1},\omega_{2}}
\end{eqnarray}
that, of course, coincides with Eq.(\ref{appa:eq17}).

For the second term (\ref{appb:eq10}), an approximation is required as the effect of three and four-body interactions is included.
Following Ref.\cite{Barker67}, we assume molecules in different intervals to be uncorrelated
\begin{eqnarray}
\label{appb:eq13}
\left \langle N_{l} N_{m} \right \rangle_{0} - \left \langle N_{l} \right \rangle_{0} \left \langle N_{m} \right \rangle_{0}&=& 0 \qquad
\qquad l \ne m \,\mbox{,}
\end{eqnarray}
and the fluctuations within a given interval, being related to the hard-spheres compressibility 
\begin{eqnarray}
\label{appb:eq14}
\left \langle N_{l}^2 \right \rangle_{0} - \left \langle N_{l} \right \rangle_{0}^2 &=& \left \langle N_{l} \right \rangle_{0} k_B T
\left(\frac{\partial \rho}{\partial P} \right)_0\,\mbox{.}
\end{eqnarray}
Substitution of Eqs.(\ref{appb:eq13}) and (\ref{appb:eq14}) into Eq.(\ref{appb:eq12}), along with Eq.(\ref{appb:eq11}), leads to
\begin{eqnarray}
\label{appb:eq15}
\beta F_{2}&=& -\frac{1}{4} k_B T \rho N \left(\frac{\partial \rho}{\partial P} \right)_0 \int d \mathbf{r}g_{0} \left(r\right) 
\left \langle \left[\beta \Phi_{\text{I}} \left(r,\Omega,\omega_{1},\omega_{2}  \right) \right]^2
\right \rangle_{\omega_{1},\omega_{2}}\,\mbox{,}
\end{eqnarray}
which is the extension of the Barker-Henderson result \cite{Barker67} to angular dependent potentials.
\section{Determination of the phase coexistence curves} 
\label{sec:appc}
To illustrate how the phase coexistence curves are found numerically, we consider in the following the phase separation into a gas and a liquid phase; the fluid-solid coexistence curve is determined correspondingly.
Our aim is to solve {Eqs.~(\ref{coexistence:eq2a})} and (\ref{coexistence:eq2b}) for the two unknown particle densities $\rho^{*}_{g}$ and $\rho^{*}_{l}$ of the gaseous and liquidus phase, respectively.
Using the common tangent construction, the concentration of the density of the gaseous and liquidus phase can be found geometrically \cite{Dijkstra99}.
In practice, however, $\rho^{*}_{g}$ and $\rho^{*}_{l}$ is determined numerically by solving {Eqs.~(\ref{coexistence:eq2a})} and (\ref{coexistence:eq2b}) simultaneously using a nonlinear root finding algorithm.
To illustrate this procedure, we rewrite {Eqs.~(\ref{coexistence:eq2a})} and (\ref{coexistence:eq2b}) as
\begin{eqnarray}
\label{h1}
h_{1}\left(\rho^{*}_{g},\rho^{*}_{l}\right) & \equiv & P_{g}^{*} \left(T^{*},\rho^{*}_{g} \right)- P_{l}^{*} \left(T^{*},\rho^{*}_{l} \right) = 0 \\
\label{h2}
h_{2}\left(\rho^{*}_{g},\rho^{*}_{l}\right) & \equiv & \mu_{g}^{*} \left(T^{*},\rho^{*}_{g} \right)- \mu_{l}^{*} \left(T^{*},\rho^{*}_{l} \right) = 0 \,\mbox{,}
\end{eqnarray} 
\noindent where we have introduced the functions $h_{1}\left(\rho^{*}_{g},\rho^{*}_{l}\right)$ and $h_{2}\left(\rho^{*}_{g},\rho^{*}_{l}\right)$. 
Since $T^{*}$ is kept fixed in the following, we have written $h_{1}$ and $h_{2}$ as function of $\rho^{*}_{g}$ and $\rho^{*}_{l}$ only.
By defining $\vec{x}=(\rho^{*}_{g},\rho^{*}_{l})^{t}$ and $\vec{h}=(h_{1},h_{2})^{t}$, where the subscript $t$ denotes the transposed matrix, our task of finding the concentrations of the two coexisting phases at constant $T^{*}$ is expressed in the form,
\begin{equation}
\label{h}
\vec{h}\left(\vec{x}\right) = 0\,\mbox{.}      
\end{equation}
\noindent This set of two nonlinear integral equation with two unknown variables is solved by using a well-tested implementation of the Newton-Raphson method \cite{Press92}, which solves {Eq.~(\ref{h})} iteratively as briefly described in the following.
First, a start value $\vec{x}_{0}$ is chosen, and the gradient $\nabla\vec{h}(\vec{x}_{0})$ is calculated.
The new value $\vec{x}_{1}$ is found by a downhill step,       
\begin{equation}
\label{newton_step}
\vec{x}_{1} = \vec{x}_{0} - J^{-1}\vec{h}(\vec{x}_{0})\,\mbox{.} 
\end{equation}
\noindent Here, $J$ is the Jacobian matrix which incorporates the partial derivatives of $h_{1}$ and $h_{2}$. 
This step is repeated, $\vec{x}_{1}$ $\rightarrow$ $\vec{x}_{2}$ $\rightarrow$ $\vec{x}_{3}$ $\rightarrow$ $\ldots$, until the fix point $\vec{x}_{n}=\vec{x}^{*}$ with  
\begin{equation}
\label{h_fixpoint}
\vec{h}\left(\vec{x}^{*}\right) = \vec{0}\,\mbox{,}      
\end{equation}  
\noindent is found.    
It is important to note here that the root finding procedure requires the evaluation of $\vec{h}(\vec{x})
$ at discrete points $\vec{x}_{i}$ only.
The nonlinear solver just steps down $\vec{h}(\vec{x})$ until {Eq.~(\ref{h})} is fulfilled to a prescribed threshold. 
Since the evaluation of $\vec{h}(\vec{x})$ at $\vec{x}=\vec{x}_{i}$ demands the calculation of several integrals, see {Eqs.(\ref{perturbation:eq5})} and (\ref{perturbation:eq9}), $\vec{h}(\vec{x})$ cannot be expressed in an analytical form.
Hence, the nonlinear solver calls a subroutine which calculates both the free energy and its gradient for each iteration step $\vec{x}_{i}$.
The free energy is evaluated using the Chebyshev quadrature and the derivatives in {Eq.~(\ref{newton_step})} are calculated using Ridder's implementation of Neville's algorithm \cite{Press92}.

After having found the two coexisting densities $\rho^{*}_{g}$ and $\rho^{*}_{l}$ at a given $T^{*}$, this procedure is repeated for a set of temperatures to map out the gas-liquid coexisting curve. 
The fluid-solid curve is calculated in exactly the same manner by equating the chemical potential and the pressure of the fluid and solid phase, {Eqs.~(\ref{coexistence:eq4a})} and (\ref{coexistence:eq4b}), respectively.


\bibliographystyle{apsrev}

\clearpage
\clearpage
\clearpage
\clearpage
\clearpage
\clearpage
\clearpage
\end{document}